\documentclass[runningheads]{llncs}
\usepackage[T1]{fontenc}
\usepackage{graphicx}
\usepackage{listings}
\usepackage{makecell}

\begin{document}
\title{Exploring Fast Fourier Transforms on the Tenstorrent Wormhole}
%
%
\author{Nick Brown\inst{1}\orcidID{0000-0003-2925-7275} \and
Jake Davies\inst{1} \and
Felix Le Clair\inst{2}}
\authorrunning{N. Brown et al.}
%
\institute{EPCC, Bayes Centre, 47 Potterrow, Edinburgh, UK \and
Tenstorrent, 2600 Great America Way, Santa Clara, California, USA}
\maketitle              
\begin{abstract}
Whilst numerous areas of computing have adopted the RISC-V Instruction Set Architecture (ISA) wholesale in recent years, it is yet to become widespread in HPC. RISC-V accelerators offer a compelling option where the HPC community can benefit from the specialisation offered by the open nature of the standard but without the extensive ecosystem changes required when adopting RISC-V CPUs.
In this paper we explore porting the Cooley–Tukey Fast Fourier Transform (FFT) algorithm to the Tenstorrent Wormhole PCIe RISC-V based accelerator. Built upon Tenstorrent's Tensix architecture, this technology decouples the movement of data from compute, potentially offering increased control to the programmer. Exploring different optimisation techniques to address the bottlenecks inherent in data movement, we demonstrate that for a 2D FFT whilst the Wormhole n300 is slower than a server-grade 24-core Xeon Platinum CPU, the Wormhole draws around 8 times less power and consumes around 2.8 times less energy than the CPU when computing the Fourier transform.

\end{abstract}
\section{Introduction}

The recent availability of high core count commodity available RISC-V CPUs \cite{brown2025performance} is driving increased interest in the role of RISC-V for HPC \cite{brown2023risc}. However, there is still some way to go for the ecosystem to fully support CPU based RISC-V supercomputers, and instead a more gradual short term adoption route is likely to be in RISC-V based PCIe accelerator cards. The major benefit of these is that they fit into existing, x86 or AArch64, systems as an add-ons. Several vendors are developing such accelerator cards, often aimed at Artificial Intelligence (AI) and Machine Learning (ML) workloads, driven by the current boom in AI. Indeed, each vendor is taking a different approach to the design of their technology based upon a set of principals that they consider important, and this demonstrates the flexibility provided by RISC-V where hardware designers can use the standard in the manner most suitable to them.

Whether initially designed for ML or HPC, RISC-V accelerator hardware fundamentally provides the building blocks to accelerate mathematical operations. Consequently, there is a role for these accelerator technologies to be leveraged by the HPC community, and one such RISC-V based accelerator card is the Wormhole developed by Tenstorrent. Available for purchase at a modest price, this commodity card is, as of 2025, one of the few RISC-V based accelerators that are publicly available. The availability and moderate price-point not only means that these can be leveraged in best-of-class supercomputers, but furthermore that they are also suitable for smaller HPC machines and even workstations. Indeed, Tenstorrent have opened up their entire software stack and work in the open, in collaboration with the wider community.

In this paper we calculate the Discrete Fourier Transform (DFT) by porting the Cooley–Tukey Fast Fourier Transform (FFT) algorithm to the Tenstorrent Wormhole accelerator. An algorithm that is ubiquitous, not only in scientific computing but also far more generally such as signal processing, our hypothesis was that the decoupling of data movement from compute provided by the Tenstorrent architecture has the potential to deliver performance and energy efficiency improvements compared to running on a CPU. The paper is structured as follows; in Section \ref{sec:bg} we survey the Tenstorrent Tensix architecture and describe the FFT algorithm, before reporting the experimental setup in Section \ref{sec:experimental_setup} used throughout this paper. Section \ref{sec:wormhole-fft} describes the design of the FFT algorithm for the Tenstorrent architecture and then explores the efficacy of optimisation techniques that aim at reducing the overhead of data movement. We then compare all 24-cores of a Xeon Platinum Cascade Lake CPU against 64 Tensix cores on the n300 for a 2D FFT in Section \ref{sec:2d-fft}, before drawing conclusions and discussing further work in Section \ref{sec:conc}.

The novel contributions of this paper are:

\begin{itemize}
    \item We undertake, to the best of our knowledge, the first study of porting the FFT algorithm to a RISC-V PCIe accelerator.
    \item An exploration of optimisation strategies for data reordering on the Tenstorrent Tensix architecture.
    \item A performance and energy efficiency analysis of an FFT solver on the Tenstorrent Wormhole against a server grade CPU.
\end{itemize}

\section{Background}
\label{sec:bg}
\subsection{The Tenstorrent architecture}

\begin{figure}[htb]
\centering
 \includegraphics[width=\columnwidth]{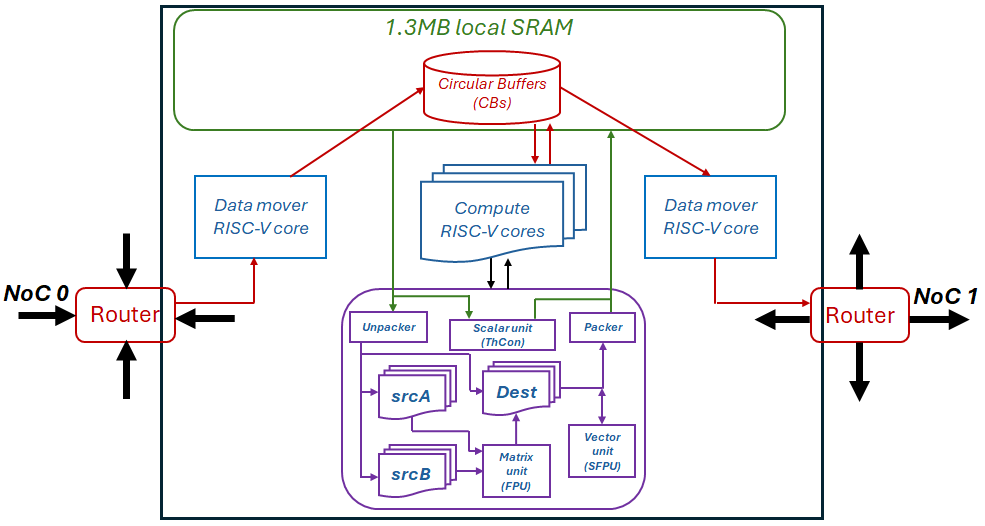}
\caption{A single Tensix core in the Wormhole accelerator, containing five RISC-V \emph{baby cores}, 1.3MB of local SRAM memory, a compute engine and two routers to the Network on Chip (NoC).}	
\label{fig:tensix}
\end{figure}

The Grayskull, Wormhole and Blackhole PCIe accelerator cards, developed by Tenstorrent, are built around the Tensix architecture. As sketched in Figure \ref{fig:tensix}, each Tensix core contains five RISC-V CPUs, known as \emph{baby cores}, 1.3MB of local SRAM, two routers each of which are connected to separate Networks on Chip (NoCs), and compute engine. Out of the five RISC-V baby cores, one of these is for moving data into the Tensix core, one for moving data out, and three drive the compute engine. The compute engine itself provides scalar (ThCon), vector (SFPU) and matrix (FPU) units that support a range of precisions up to FP32, which we use throughout this paper, although the matrix unit relaxes IEEE compliance. Out of the three RISC-V baby cores driving the compute engine, one called \emph{UNPACK} issues instructions to the unpacker in the compute engine, which copies data from SRAM into the source registers, \emph{srcA} and \emph{srcB}. The \emph{MATH} RISC-V baby compute core issues instructions to the ThCon, SFPU and FPU units, directing them to undertake operations on source registers. The third RISC-V baby compute core, \emph{PACK}, packs (or copies) result data from the \emph{dst} register to SRAM. The \emph{srcA} and \emph{srcB} registers are 4KiB in size, holding a maximum of 1024 single precision floating point numbers, and the \emph{dst} register is of size 32KiB and segmented into 16 chunks \cite{corsix}. To avoid being a bottleneck, input and result values are never transferred through the RISC-V compute baby cores, but instead instruct the unpacker and packer in the compute unit to accesses SRAM directly.

The \emph{dst} register is used as both an input to, and output from, the vector unit and this is why it is split into 16 segments. Consequently, compute can undertaken by the matrix unit and results can then be provided to the vector unit for further processing. Indeed, the matrix unit only supports a restricted set of operations including matrix multiplication, addition, subtraction, multiplication, transposition and reduction. By contrast, the vector unit is more general and in addition to binary arithmetic also provides support for operations including squares, square root, cos, tan, sin, and conditionals. In-fact, the programmer can write their own code directly for the SFPU and the compiler will then vectorise this. The matrix unit can perform 2048 floating point multiplications and 2048 floating point additions per cycle. The vector unit comprises eight registers, each with thirty two 32-bit lanes. 

The abstraction of \emph{Circular Buffers} (CBs) represents data held in local SRAM. These are First In First Out (FIFO) queues and follow a producer-consumer model, where a page of data is requested in the CB by the producer, populated and pushed to be made available. The consumer blocks on the availability of a page, and once it is, then consumes the data before freeing the memory so that this can be reused. CBs combine semantics around memory and synchronisation, enabling coordination between the RISC-V baby cores. An example is the common situation where a data mover core requests a CB page, then fills this with data read from external DRAM and pushes this. The RISC-V compute cores then consume this CB, with the \emph{UNPACK} core instructing the \emph{unpacker} in the compute engine to copy data from this page in the CB into a target register.

In this work we focus on a Wormhole n300 which contains 120 Tensix cores that are provided as two, 60-core, chips running at 1 GHz. There is a total of 24GB of GDDR6, external, memory on the board which is split across twenty four banks each of 1GB, with twelve banks directly connected to each chip \cite{n300-datasheet}. Whilst, at the time of writing, Tenstorrent have recently released their next generation accelerator, the Blackhole, the Wormhole is currently by far the more common technology and due to the same underlying architecture, lessons learnt on this generation will apply to the next.

The TT-Metalium framework, \emph{tt-metal}, is Tenstorrent’s direct programming SDK which exposes access to the hardware, providing an API for kernel development. The SDK provides an API that programmers can use to undertake a range of low level activities such as the movement of data, driving the compute engine, and interacting with Circular Buffers (CBs). The programmer develops three kernels, one for each data movement RISC-V baby core and one for the compute cores. 

Previous work \cite{brown2024accelerating} explored stencil applications on the first generation, Grayskull e150, Tenstorrent accelerator. This work demonstrated that it was possible for the architecture to perform comparatively to a server-grade CPU but at five times less energy. However, a major challenge was in reworking the algorithm to suit the architecture, with the final version of the code around 160 times faster on the Grayskull than the initial, naive, kernel implementation.

\subsection{Fast Fourier Transforms (FFTs)}

The calculation of the Discrete Fourier Transform (DFT) is of critical importance to a wide variety of applications ranging from digital signal processing to solving systems of partial differential equations. Converting between the spatial and frequency domains, the former represents a collection of values and the later describes the rate at which these values are changing. Fast Fourier Transforms (FFTs) are a class of fast algorithms that compute the DFT, with the Cooley–Tukey algorithm \cite{strang2000linear} being the most common and the focus in this work. Indeed, the FFT algorithm was described as \emph{the most important numerical algorithm of our lifetime} \cite{strang2000linear} as it reduces the computational complexity to O(N log N). Cooley-Tukey follows a divide-and-conquer approach and simple implementations adopt a recursive form, however for performance this is expressed iteratively with an outer loop of log N steps.

\begin{figure}[htb]
\centering
 \includegraphics[width=\columnwidth]{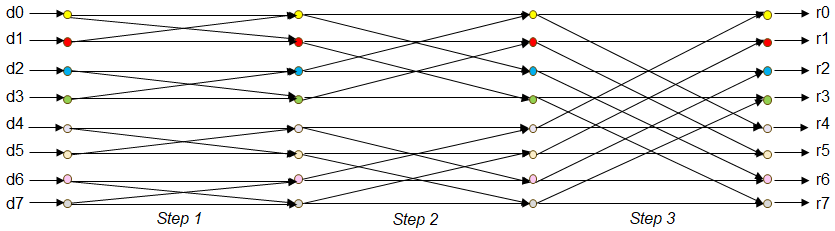}
\caption{Illustration of the steps involved and data dependencies with a radix-2 Cooley-Tukey FFT on eight values}	
\label{fig:fft}
\end{figure}

Figure \ref{fig:fft} illustrates an FFT operating on eight input values, requiring three steps. We follow a 2-radix approach, were in the first step calculations involving neighbouring values are undertaken, in the second step intermediate results are then calculated between neighbour plus (or minus) 2, and in the third step intermediates with neighbour plus (or minus) four. 

Consequently, in addition to the calculation itself, the movement of data is an important consideration as from one step to the next the compute requires different pairs of numbers. This is illustrated in Listing \ref{lst:fft} where, for each step, individual \emph{points} in groups of \emph{spectra} are processed. There are two data elements per point, and it can be seen that the indices of these, \emph{d0\_data\_index} and \emph{d1\_data\_index}, are calculated at lines 6 and 7, with \emph{matching\_second\_point} providing the offset for the second element, for instance 1, 2, 4 for steps 1, 2 and 3 in Figure \ref{fig:fft}. Values \emph{f0} and \emph{f1} are then calculated based upon twiddle factors and the second element of data at lines 9 and 10, before these values are applied to the real and imaginary components of the two data elements at lines 12 to 15. Twiddle factors are trigonometric constants that are multiplied by the data during execution of the algorithm.

\begin{lstlisting}[frame=lines, label=lst:fft, caption=Sketch of FFT algorithm which is computing using real and imaginary components]
for (int step=0; step <= num_steps; step++) {    
    int matching_second_point=...
    for (int spectra=0; spectra < num_spectra_in_step; spectra++) {
      int twiddle_index=spectra << (num_steps-step);
      for (int point=0; point < domain_size; point+=increment_next_point_in_step) {                
        int d0_data_index=(spectra + point);
        int d1_data_index=(spectra + point + matching_second_point);
        
        float f0=(data[d1_data_index].r * twiddle_factors[twiddle_index].r) - (data[d1_data_index].i * twiddle_factors[twiddle_index].i);
        float f1=(data[d1_data_index].r * twiddle_factors[twiddle_index].i) + (data[d1_data_index].i * twiddle_factors[twiddle_index].r);

        data[d1_data_index].r=data[d0_data_index].r - f0;
        data[d1_data_index].i=data[d0_data_index].i - f1;
        data[d0_data_index].r=data[d0_data_index].r + f0;
        data[d0_data_index].i=data[d0_data_index].i + f1;
      }
    }
  }
\end{lstlisting}

\section{Experimental setup}
\label{sec:experimental_setup}
Results reported from the experiments run throughout this paper are averaged over five runs. All Tenstorrent runs were undertaken on a Wormhole n300 connected to the host system by PCIe Gen 3. The host machine contains a 26-core 8170 Skylake Intel Xeon Platinum CPU and 128 GB of DRAM. All experiments are built using version 0.56 of the tt-metal framework, and Clang 17 is used to compile host codes. CPU runs are undertaken on a 24-core 8260M Cascade Lake Xeon Platinum CPU, which is equipped with 512GB of DRAM and codes are compiled using GCC version 11.2. All codes are compiled at optimisation level three. Multi-core runs on the CPU are multi threaded using OpenMP. Energy usage on the CPU is based upon values reported by RAPL, and on the n300 from the Tenstorrent System Management Interface (TT-SMI).

All results reported in this paper are running single precision, FP32, which is the maximum precision supported by the n300. Unless otherwise stated, performance numbers reported for the Wormhole are execution time only.

\section{Porting Fourier transforms to the Wormhole}
\label{sec:wormhole-fft}
In this section we focus on porting FFTs to the Wormhole and optimising for a single Tensix core. Whilst complex numbers are not directly supported by the Tensix's compute engine, it is possible to work with the real and imaginary components individually as illustrated in Listing \ref{lst:fft} and this is the approach adopted here.

\begin{figure}[htb]
\centering
 \includegraphics[width=\columnwidth]{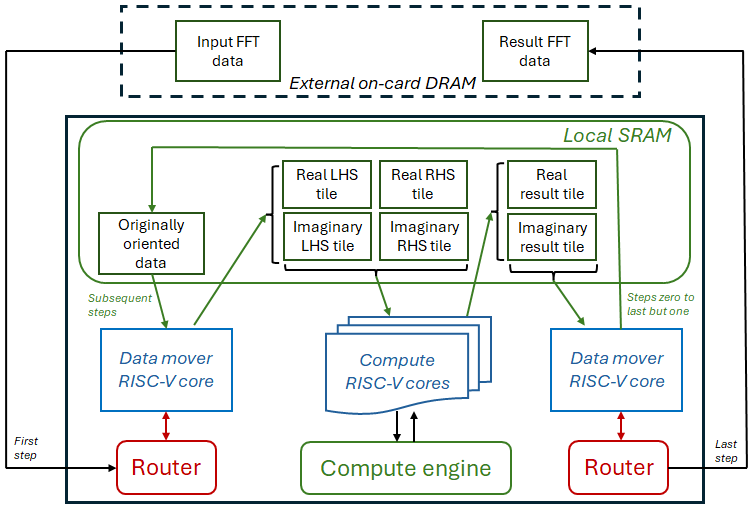}
\caption{Basic design approach of running FFTs on a Tensix core}	
\label{fig:fft-tensix}
\end{figure}

Figure \ref{fig:fft-tensix} illustrates the design that we initially leveraged when porting the FFT algorithm to the Tensix core. For each step, the in data mover core will read the input data for that step. This is read either from external on-card DDR for the first step or from local SRAM for subsequent ones, and a page in four CBs is populated with the data correctly ordered for that specific step. These CBs are the computation's Left Hand Side (LHS) and Right Hand Side (RHS) for both real and imaginary components. These are then used as an input to the compute core which provides these to the compute unit to undertake the calculations. The output of these calculations is stored in real and imaginary CBs which are consumed by the out data mover core which then reorders the data into the original order and will either store this in SRAM, ready to be consumed by the next step, or external DDR for the final results. Consequently, for each step we undertake two reorderings of data, from the original data order to the order required for pair wise operations by that current step, and then results are ordered from this stepwise ordering back to the original orientation. 

Figure \ref{fig:fft-tensix} is simplified somewhat, for instance twiddle factors are calculated by the compute engine on initialisation and stored in SRAM but these do not change from step to step. Listing \ref{lst:fft_tt} sketches the FFT compute kernel, at line 1 operating in steps, and lines 2 and 3 waiting for pages of real and imaginary data to be available in these CBs for the second element of data. Lines 5 to 13 then use the SFPU to compute \emph{f0} and \emph{f1} before the real and imaginary pages of the first element of data in the CBs are waited on at lines 15 and 16. Lines 21 to 28 then applying \emph{f0} and \emph{f1} to the real and imaginary components of the two data values before popping the input data CBs at lines 27 to 30, which will enable the pages of memory to be reused by the next step. The \emph{cb\_int0} and \emph{cb\_int1} CBs are circular buffers used to hold intermediate calculation data, and the \emph{cb\_twiddle\_r} and \emph{cb\_twiddle\_i} hold twiddle values.

\begin{lstlisting}[frame=lines, label=lst:fft_tt, caption=Sketch of FFT compute kernel code]
for (int step=0; step <= num_steps; step++) {
  cb_wait_front(cb_data1_r, 1);
  cb_wait_front(cb_data1_i, 1);

  // Calculate f0
  maths_sfpu_op<MUL>(cb_data1_r, cb_twiddle_r, cb_int0);
  maths_sfpu_op<MUL>(cb_data1_i, cb_twiddle_i, cb_int1);
  maths_sfpu_op<SUB, true, true>(cb_int0, cb_int1, cb_f0);

  // Calculate f1    
  maths_sfpu_op<MUL>(cb_data1_r, cb_twiddle_i, cb_int0);
  maths_sfpu_op<MUL>(cb_data1_i, cb_twiddle_r, cb_int1);
  maths_sfpu_op<ADD, true, true>(cb_int0, cb_int1, cb_f1);

  cb_wait_front(cb_data0_r, 1);
  cb_wait_front(cb_data0_i, 1);

  // Calculate data_1 real
  maths_sfpu_op<SUB>(cb_data0_r, cb_f0, cb_out_data1_r);
  // Calculate data_1 imaginary
  maths_sfpu_op<SUB>(cb_data0_i, cb_f1, cb_out_data1_i);
  // Calculate data_0 real
  maths_sfpu_op<ADD>(cb_data0_r, cb_f0, cb_out_data0_r);
  // Calculate data_0 imaginary
  maths_sfpu_op<ADD>(cb_data0_i, cb_f1, cb_out_data0_i);
   
  cb_pop_front(cb_data0_r, 1);
  cb_pop_front(cb_data0_i, 1);
  cb_pop_front(cb_data1_r, 1);
  cb_pop_front(cb_data1_i, 1);
}
\end{lstlisting}

The \emph{maths\_sfpu\_op} function that is called in Listing \ref{lst:fft_tt} is sketched in Listing \ref{lst:sfpu_fn}. We developed this as a helper function due to the boilerplate code that is required for each operation, where a lock on the \emph{dst} register must first be acquired by the \emph{MATH} RISC-V compute core, tiles are then copied in from page location 0 in CBs \emph{cb\_in\_1} and \emph{cb\_in\_2} to segments 0 and 1 of the \emph{dst} register at lines 7 and 8, before the compute operation is performed on \emph{dst} register segments 0 and 1 at lines 9 to 17 and results written to the 0 segment. The \emph{MATH} RISC-V compute core then releases its lock on the \emph{dst} register at line 18, before a page in the \emph{cb\_tgt} CB is waited on at line 22, and then the \emph{PACK} RISC-V compute core acquires a lock on the \emph{dst} register, packs values from the 0 segment of the \emph{dst} register to a page in the CB, and releases the lock at lines 23 to 25, before making the page in \emph{cb\_tgt} CB available. Optionally, this function will also wait for pages in input CBs and free these up, and it can be see in Listing \ref{lst:fft_tt} that this is used when working with intermediate values. Whilst we focus here on the vector unit, SFPU, we also provide versions of this maths operation function for computing with matrix unit, FPU, which is similar to Listing \ref{lst:sfpu_fn} but somewhat simplified because the Metalium API call performs copying of input data into the \emph{srcA} and \emph{srcB} registers itself. 

\begin{lstlisting}[frame=lines, label=lst:sfpu_fn, caption=Sketch of SFPU maths helper function]
template <int OPERATION, bool CB_OP_IN1=false, bool CB_OP_IN2=false>
void maths_sfpu_op(uint32_t cb_in_1, uint32_t cb_in_2, uint32_t cb_tgt) {
    // Copy from input CBs into index 0 and 1 of DST regs
    if constexpr(CB_OP_IN1) cb_wait_front(cb_in_1, 1);
    if constexpr(CB_OP_IN2) cb_wait_front(cb_in_2, 1);
    tile_regs_acquire();
    copy_tile(cb_in_1, 0, 0);
    copy_tile(cb_in_2, 0, 1);
    if (OPERATION == ADD) {        
        add_binary_tile(0, 1);
    } else if (OPERATION == SUB) {
        sub_binary_tile(0, 1);
    } else if (OPERATION == MUL) {
       mul_binary_tile(0, 1);
    } else if (OPERATION == DIV) {
       div_binary_tile(0, 1);
    }
    tile_regs_commit(); 
    if constexpr(CB_OP_IN1) cb_pop_front(cb_in_1, 1);
    if constexpr(CB_OP_IN2) cb_pop_front(cb_in_2, 1);
    // Result is in 0 index of DST regs, extract
    cb_reserve_back(cb_tgt, 1);
    tile_regs_wait();
    pack_tile(0, cb_tgt);
    tile_regs_release();
    cb_push_back(cb_tgt, 1);
}
\end{lstlisting}

Table \ref{tab:initial-perf} reports performance of an FFT implementation for a problem size of 16384 elements on a single Xeon Platinum (Cascade Lake) CPU core, along with different versions on a single Tensix core in the Wormhole. The first version on the Wormhole is as illustrated in Figure \ref{fig:fft-tensix} and Listing \ref{lst:fft_tt}, using the SFPU for the calculation. It can be seen that it is around eight times slower than a single core is on the Xeon Platinum CPU.

\begin{table}[htb]
    \centering
    \caption{Runtime on Tenstorrent Tensix and CPU core executing FFT algorithm for problem size 16384 of elements.}
    \label{tab:initial-perf}
    \begin{tabular}{|c|c|}
    \hline           
     \textbf{Version} & \textbf{Runtime (ms)} \\
      \hline
    Xeon Platinum CPU single core & 1.85 \\
    \hline
    Initial & 14.39\\
    Chunked & 9.38\\
    Data copy by ThCon & 7.56\\
    128-bit copies & 6.61\\
    Single data copy & 5.31\\
    \hline
    \end{tabular}
\end{table}

To explore the underlying reasons for this performance on the Tensix we experimented with disabling certain components in the algorithm and the performance that these individual components delivered is reported in Table \ref{tab:strip-perf}. The major overhead is in the reordering of data, for instance when disabling data reordering for reading the runtime halves, and when also disabling write reordering the runtime is around a sixteenth of the original. Clearly data reordering is a significant bottleneck, and by comparing the performance when only read or write reordering is enabled in Table \ref{tab:strip-perf} it can be observed that read reordering is the more expensive of the two operations. Incidently, compute-only performance was comparable regardless of whether we used the FPU (matrix unit) or SFPU (vector unit) to undertake mathematical operations.

\begin{table}[htb]
    \centering
    \caption{Performance on a single Tensix core executing FFT algorithm for 16384 elements when specific components are enabled (Y) or disabled (N).}
    \label{tab:strip-perf}
    \begin{tabular}{|ccccc|c|}
    \hline           
     \makecell{\textbf{External} \\ \textbf{read}} & \makecell{\textbf{Read} \\ \textbf{reorder}} & \textbf{Compute} & \makecell{\textbf{Write} \\ \textbf{reorder}} & \makecell{\textbf{External} \\ \textbf{write}} & \makecell{\textbf{Runtime} \\ \textbf{(ms)}} \\
      \hline   
    Y & Y & Y & Y & Y & 14.4\\
    Y & N & Y & Y & Y & 7.3\\
    N & N & Y & Y & Y & 7.3\\
    N & Y & Y & N & N & 10.5\\
    Y & Y & Y & N & N & 10.6\\
    N & N & Y & N & Y & 0.9\\
    N & N & Y & N & N & 0.9\\                
    \hline
    \end{tabular}
\end{table}

Our initial approach placed the entire domain into single CB pages step by step, for instance loading and reordering all the data, then computing with this, and lastly reordering and writing this all out. Whilst placing the entire domain in a single page was the simplest approach, it meant that different components of the Tensix core could not run concurrently, for instance whilst data loading was occurring by the data mover RISC-V core, the compute engine and data writing kernels were idle. Consequently, we enhanced the code to operate in chunks, \emph{chunked} in Table \ref{tab:initial-perf}, where the entire domain is split into segments and consequently the components of the Tensix core can run in parallel on different chunks. 

When reordering the data, this data is loaded into the data mover cores from SRAM and then written out to a new location. The cores themselves are fairly limited and designed more for driving components such as the routers, compute engine and marshalling control with the CBs rather than undertaking extensive data loads and stores. Instead, the compute engine contains a scalar unit, ThCon, which itself can load and write data and the hypothesis \cite{corsix} was that this could reduce the runtime due to improved performance of compute engine to SRAM data transfers. Whilst there are no API calls to ThCon exposed directly by Metalium, it is possible to program this via intrinsics provided by the Tenstorrent Low Level Kernels (llk) underlying library. The code for this is sketched in Listing \ref{lst:llk_thcon} which loads 32-bit data from SRAM. The address \emph{from\_addr} is decomposed into a base address and offset and then stored into registers \emph{0} and \emph{1} via the \emph{TT\_SETDMAREG} intrinsic. These registers are then provided to the \emph{TT\_LOADIND} intrinsic, with register \emph{2} provided as the destination to hold the loaded value. There are four tiles of data being reordered, the LHS and RHS for both real and imaginary components, and we found the most effective approach was to distribute these across all three RISC-V compute cores; \emph{UNPACK}, \emph{MATH} and \emph{PACK} which individually issue instructions to ThCon for their respective pages.

\begin{lstlisting}[frame=lines, label=lst:llk_thcon, caption=Sketch of loading data from SRAM via ThCon using LLK intrinsics]
uint32_t base_addr=from_addr / 16;
uint32_t addr_offset=from_addr-(base_address*16);

TT_SETDMAREG(0,LOWER_HALFWORD(addr_offset),0,LO_16(0));
TT_SETDMAREG(0,UPPER_HALFWORD(addr_offset),0,HI_16(0));
TT_SETDMAREG(0,LOWER_HALFWORD(base_addr),0,LO_16(1));
TT_SETDMAREG(0,UPPER_HALFWORD(base_addr),0,HI_16(1));

TT_LOADIND(p_ind::LD_32bit, LO_16(0), p_ind::INC_4B, 2, 1);
\end{lstlisting}

The result of this optimisation is reported as \emph{Data copy by ThCon} in Table \ref{tab:initial-perf} where it can be seen that it reduced the runtime by around a millisecond. It is possible for ThCon to read and write data of size 8, 16, 32 or 128 bits, and due to us working with FP32 up until this point all data accesses were 32 bit. However, when the kernel reorders input data, stores are all contiguous and when reordering results for writing the loads are all contiguous. The kernel was therefore modified to unroll the reordering loop by four and to use 128-bit wide data accesses for contiguous data. This is reported as \emph{128-bit copies} in Table \ref{tab:initial-perf} and it can be seen that this further reduced the runtime. 

\begin{figure}[htb]
\centering
 \includegraphics[width=\columnwidth]{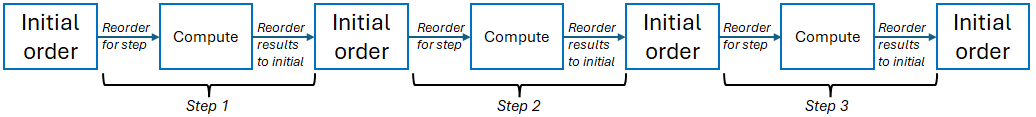}
\caption{Two data reordering stages per step, converting back to the original data order between steps}	
\label{fig:two-data-reordering}
\end{figure}

\begin{figure}[htb]
\centering
 \includegraphics[width=\columnwidth]{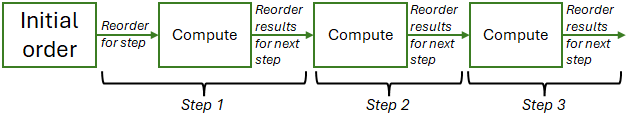}
\caption{One data reordering stage per step, converting to the next step's data order directly to reduce the overall number of expensive data reorderings required.}	
\label{fig:one-data-reordering}
\end{figure}

To this point, as per Figure \ref{fig:two-data-reordering}, there were two data reorderings per step. This is because, at the end of a step, the data is reordered back to the original data order. Whilst this was the simplest approach from a code perspective, the reordering of data is expensive and we therefore modified this to instead reorder data to the arrangement required by the next step. This is illustrated in Figure \ref{fig:one-data-reordering}, and it reduces the number of reorderings per step to one, apart from the initial and last step. However, this increases the complexity of the code which resulted in link errors where the compiler reported \emph{ \'.bss\' will not fit in region \`TRISC2\_LOCAL\_DATA\_MEM\'}. The bss section holds uninitialized data and this was overflowing, likely due to too many variables. To address this we edited the \emph{kernel\_trisc2.ld} linker script to increase the \emph{TRISC2\_LOCAL\_DATA\_MEM} from 1280 to 3328 bytes.

Performance of the one-copy optimisation is reported as \emph{Single data copy} in Table \ref{tab:initial-perf} and, whilst it resulted in a performance improvement, the limited nature of this surprised us as we had expected a more considerable reduction in runtime. When investigating we found that this was due to all data accesses now being non-contiguous and therefore 32 rather than 128 bits wide. Effectively we had increased the number of individual memory accesses with a larger number of narrow accesses, whereas before we had fewer wider accesses.

\section{Scaling up with a 2D FFT}
\label{sec:2d-fft}
\begin{figure}[htb]
\centering
 \includegraphics[width=0.6\columnwidth]{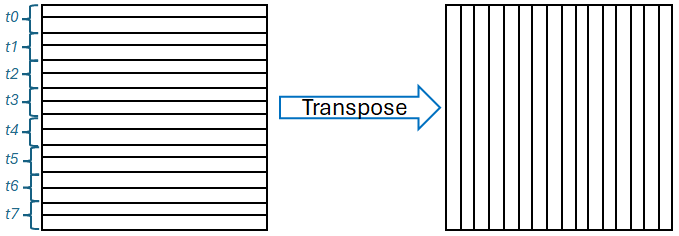}
\caption{Illustration of data movement involved in a 2D FFT}	
\label{fig:2dfft}
\end{figure}

Thus far we have concentrated on running over a single Tensix core and in this section we scale up by studying a 2D FFT. Figure \ref{fig:2dfft} illustrates the data orientation required by such an operation, where rows of data are distributed across the Tensix cores, here \emph{t0} to \emph{t7}, and one dimensional FFTs are first executed across each row. The data is then globally transposed so that each Tensix core receives a column of data comprising parts made up from across each Tensix core. Each core then undertakes another FFT on its newly transposed local data. Ultimately, an FFT has been performed across each global row and down each global column to produce the overall result. 

We enhanced our FFT implementation to work across a several locally held rows and leveraged the \emph{transpose} routine from Tenstorrent's tt-nn library to undertake transposition across the Tensix cores. For simplicity we assume an even decomposition of rows across Tensix cores, and with a problem size of 1024 by 1024 this limits us to 64 Tensix cores, each holding 16 local rows. Table \ref{tab:2dfft-perf} reports performance and power of a 2D FFT of size 1024 by 1024 elements parallelised via OpenMP on all 24 cores of the Xeon Platinum Cascade Lake CPU against 64 Tensix cores of the n300. It can be seen that whilst the CPU is faster than the n300, it draws over eight times more power, ultimately resulting in the n300 being 3.6 times more energy efficient. 

\begin{table}[htb]
    \centering
    \caption{Runtime of Wormhole n300 and entire Xeon Platinum Cascade Lake CPU executing 2D FFT for problem size 1024 by 1024 elements.}
    \label{tab:2dfft-perf}
    \begin{tabular}{|c|c|ccc|}
    \hline           
     \textbf{Version} & \makecell{\textbf{Number of} \\ \textbf{cores}} & \makecell{\textbf{Runtime} \\ \textbf{(ms)}} & \makecell{\textbf{Average Power} \\ \textbf{(Watts)}} & \makecell{\textbf{Energy usage} \\ \textbf{(J)}} \\
      \hline
    Xeon Platinum CPU & 24 & 10.24 & 353 & 3.62 \\
    Wormhole n300 & 64 & 23.56 & 42 & 0.99 \\
    \hline
    \end{tabular}
\end{table}

\section{Conclusions and future work}
\label{sec:conc}
In this paper we have explored calculating the DFT by porting the Cooley-Tukey FFT algorithm to the Tensix architecture. We explored our general design approach of building up tiles of real and imaginary data which are then used as inputs to the compute engine, and abstracting the boilerplate around Metalium mathematical operators via C++ templated functions. We found data reordering was a significant bottleneck and explored optimisation techniques to address this, with the use of ThCon and 128-bit wide memory accesses especially effective.

Whilst the performance of a single Tensix core was around 2.8 times lower than a Xeon Platinum Cascade Lake CPU core, the Wormhole n300 contains many more energy efficient Tensix cores. Consequently we scaled up our approach and explored 2D FFT demonstrating that, whilst 64 Tensix cores in the n300 were slower than the Xeon Platinum CPU, they also drew around eight times less power, ultimately with the Wormhole being 3.6 times more energy efficient.

This paper has highlighted that there is great potential for Tenstorrent technology to benefit HPC workloads, especially due to its energy efficiency nature. However, given that these accelerators have been designed for ML workloads a challenge is to understand how to adapt the algorithms and tooling to best suit general purpose HPC. To this end, more integrated control of ThCon in Metalium would be beneficial, especially if this better abstracted the reordering of data. Furthermore, the ability to load in, and store, data from the matrix and vector registers based upon a user defined mapping directly would likely improve performance as it would avoid additional data copying.

In this paper we limited ourselves to holding the entirety of the domain in local SRAM and, with other data structures required for reordering and tiling, this resulted in a maximum problem size of 16384 FP32 elements. When scaling to 2D, our problem size was 1 million elements, and this is also small. The next step will be to support larger domains by reordering from external, on-card, DRAM, using SRAM as a temporary staging area. Furthermore, we might be able to obtain improved performance within a single Tensix core by reworking the one-copy data reordering scheme to work across contiguous memory, thus resulting in single 128-bit rather than four 32-bit memory accesses. Moreover, expanding the transposition to support an uneven distribution of rows across Tensix cores will enable us to leverage all 120 cores on the n300, likely closing the gap significantly with the Xeon Platinum CPU. Whilst in this paper we focussed on a single Wormhole card, a major bottleneck for multi-dimensional FFTs is the all-to-all communications required during transposition. This work acts as a foundation that can be built upon in the future to explore how the high performance network connecting multiple n300 cards can remove this bottleneck.

\section*{Acknowledgement}
The CPU runs in this paper ran on NextGenIO which which funding from the EU Horizon 2020 research and innovation programme under grant agreement No 671591. This research was supported by an RSE personal research fellowship. We thank Tenstorrent, especially Pete Cawley, for their technical assistance. For the purpose of open access, the author has applied a Creative Commons Attribution (CC BY) licence to any Author Accepted Manuscript version arising from this submission.

%
%
%
\bibliographystyle{splncs04}
\bibliography{references.bib}

\end{document}